# Strategies for spectroscopy on Extremely Large Telescopes: III – Remapping switched fibre systems

C.L. Poppett, J.R. Allington-Smith[*] and G.J. Murray

Centre for Advanced Instrumentation, Physics Department,

Durham University, South Rd, Durham DH1 3LE

**Abstract**

We explore the use of remapping techniques to improve the efficiency of highly-multiplexed fibre systems for astronomical spectroscopy. This is particularly important for the implementation of Diverse Field Spectroscopy (DFS, described in Paper II) using highly-multiplexed monolithic fibre systems (MFS). Diverse Field Spectroscopy allows arbitrary distributions of target regions to be addressed to optimise observing efficiency when observing complex, clumpy structures such as protoclusters which will be increasingly accessible to Extremely Large Telescopes (ELTS). We show how the adoption of various types of remapping between the input and output of a Monolithic Fibre Systems can allow contiguous regions of spatial elements to be selected using only simple switch arrays. Finally we show how this compares in efficiency with integral-field and multiobject spectroscopy by simulations using artificial and real catalogues of objects. With the adoption of these mapping strategies, DFS outperforms other techniques when addressing a range of realistic target distributions. These techniques are also applicable to bio-medical science and were in fact inspired by it.

## 1 Introduction

This is the third part of a study of strategies to produce instruments for ELTs that are both affordable and efficient before the full benefits of Adaptive Optics (AO) delivering near-diffraction-limited images over a wide field become available.

The first paper (Allington-Smith, 2007; Paper 1) discussed the potential for reducing instrument size through the use of image-slicing and the evolution towards integral field spectroscopy (IFS) as the AO system delivers improved image quality. Paper II introduced the concept of Diverse Field Spectroscopy – an arbitrary combination of IFS and multiobject spectroscopy (MOS) – as relevant to studies of high-redshift proto-galactic structures as well as crowded stellar fields. It went on to derive an appropriate specification for a conceptual instrument: the *Celestial Selector* and showed how this could be implemented using Monolithic Fibre Systems - large bundles of fibres fused together in a monolithic piece - in conjunction with arrays of optical switches.

DFS, like IFS, provides complete sampling of contiguous fields at the Nyquist-Shannon sampling limit. In this paper, we explore mapping strategies between the input and output of the Monolithic Fibre Systems that maximise the systems ability to select contiguous groups of spatial elements (spaxels) without the need for highly-multiplexed switch arrays. Thus the switching layer could be made up of a large array of replicated switches, each with modest multiplex capability. By using relicable components, the construction of a highly-multiplexed spectrograph can be industrialised using the principles of production engineering already long-established outside astronomy. This should lead to more reliable, cheaper and more capable instruments.

---

[*] Email address: j.r.allington-smith@durham.ac.uk

Consideration of the example application discussed in Paper II suggests that a requirement for a full, but relatively unambitious, implementation is a system that downselects from $N \sim 10^5$ to $M \sim 10^4$ spaxels, i.e. by a factor in the general range 10-100. However there are also non-astronomical applications, and it was in fact a bio-medical application that prompted the use of remapping.

Diverse Field Spectroscopy is directly applicable to bio-medical studies of transient phenomena in *in vivo* studies of living organisms. Here it is important that all spectral and spatial data be collected simultaneously. This rules out many hyperspectral techniques already used in bio-medical science: those that use small numbers of detector pixels to repeatedly scan the source in order to build-up the datacube. As with astronomical applications, there will be a downselection because it is not possible to record the spectrum from each spatial sample at the same time.

Consideration of the bio-medical applications suggests a suitable format is $N = 2500$, $M = 100$. This provides the clinician with up to 100 regions of interest whose spectra may be monitored for the duration of the experiment or analysis. These ROIs must be selectable by the operator with as much freedom as possible – if necessary in one contiguous group or a diverse set of single spaxels - exactly as in Diverse Field Spectroscopy. The choice of $M$ is designed to be compatible with commercial slit-based spectrometers used as a standard accessory for existing microscopes.

The paper is laid out as follows. The case for remapping is made in §2 and different methods presented in §3. In §4 we present results obtained from simulations of clustered fields with clustered regions of interest and compare this with the success of IFS and MOS. Finally, in §5 we provide a reality check by applying the technique to a range of real target distributions.

# 2  Remapping

## 2.1  The advantage

The Celestial Selector concept presented in paper II may be conveniently implemented with a Monolithic Fibre Systems followed by a layer of optical switches that route selected fibres to a number of spectrographs. Ideally the switch array allow $N$ spaxels to be down-selected to $M$ ($< N$) but this is unfeasible on the scales needed where $N \sim 10^5$ and $M \sim 10^4$. The capabilities of optical switches in the telecommunication industry is generally $10 < N,M < 100$. Therefore the inputs must be divided between many switches (hereafter we shall use the generic term "cell" in preference to the implementation-specific term "switch"). One example of this was the Spatial Light Modulator array discussed in Paper II where $(n,m) = (25,1)$ where $n$ and $m$ are the number of inputs and output per cell. But dividing the full array into cells inevitably restricts the capability to select contiguous groups of spaxels. In the example given, the maximum number of spaxels in a contiguous group of 25 that could be routed to a spectrograph is only 1, and 2 for 50 etc. Adopting a system, with the same downselect factor but larger format, say (100,4) does not really help since only 4 contiguous spaxels may be addressed. However for $n > 4$, the larger formats are better since switch units with only $m = 1$ output cannot address more than 4 spaxels in a group assuming regular square tessellation of the sub-fields of each switch and only if the pointing is adjusted to put the large group on the corner of 4 adjacent cells (an example is given later).

However if the mapping between the field and the switch array is *incoherent* – i.e. randomising or re-ordering in other ways, there will be a finite chance that two or more spaxels which are adjacent in the field will be mapped to a spectrograph since they now occupy different cells (Figure 1).Thus the probability of mapping a desired selection of contiguous spaxels, or, more generally, a specific constellation of regions of interest, can be boosted so that, through judicious design by the investigator the scientific return can be increased by allowing more RoIs to be targeted in a single exposure. With *coherent* mapping, the probability would be negligible.

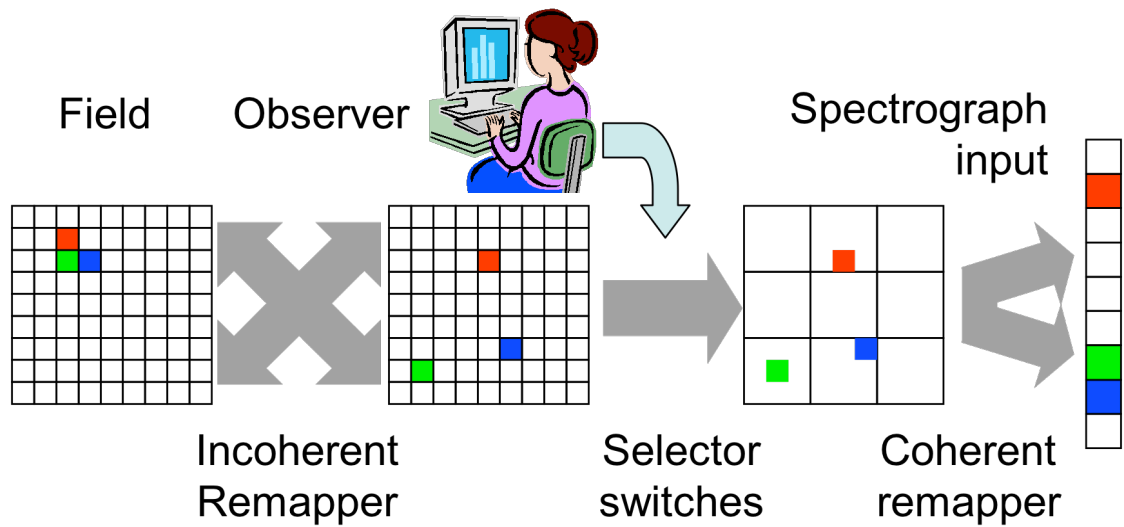

**Figure 1: Illustration of the principle of remapping applied to a downselection system.**

The division of the input fibre bundle (IFB) into tiles is for purely practical reasons, so that it can be built up from standardised units made on a production line. However it does have an important consequence when incoherent mapping is used. The incoherence can only apply within a tile: it will not be possible to incoherently map between tiles. This means that if a distribution of RoIs is defined by a nearly contiguous group of similar size to a tile, the success rate in routing spaxels in the field to the spectrographs is given simply by the ratio $m/n$. Thus the tiling imposes an upper limit on the scale size of RoIs. This limitation may be avoided if hierarchical switching is used (see §4 of Paper II).

The geometry of a fibre-based implementation (e.g. based on the MAIFU technology presented in Paper II) is defined in Paper II, but is summarised here for clarity in Figure 2.

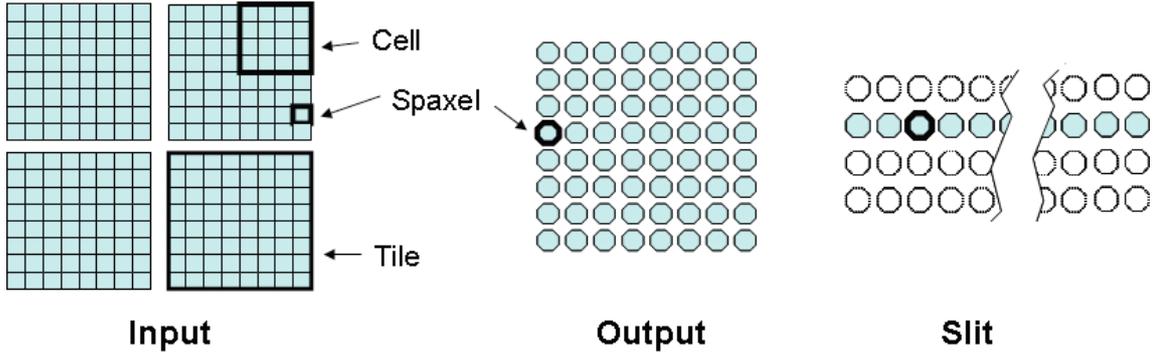

**Figure 2: Fibre system geometry. The field is divided into $N_T$ tiles each containing $N_C$ cells of $N_S$ spaxels. Within each cell, $N_D$ spaxels can be routed to the output (relating to the discussion in the text, $N_S \equiv n$ and $N_D \equiv m$). Therefore the total number of inputs and outputs respectively are defined by $N=N_T N_C N_S$, $M=N_T N_C N_D$ and the downselection factor is $G=N/M = N_S/N_D$. In this example $N_T = 4$ tiles each with $N_C = 4$ cells containing $N_S = 16$ spaxels. Of these $N = 256$ spaxels. $N_D = 4$ can be routed to the output in each cell so there are $M = 64$ possible outputs.**

## 3  Remapping schemes

### 3.1  Description

Various remapping schemes are possible. In each case we will demonstrate the principle using a schematic that shows the input and output spaxel distributions for a simple illustrative distribution of RoIs (Figure 3) and the pattern at the output of the IFB. The success of the mapping in routing target spaxels to the slit is given for each example but refers only to one realisation of the target distribution. This is supplemented in §4 by a more rigorous analysis of the probability that an arbitrary arrangement of RoIs ("constellation") can be routed to the slit.

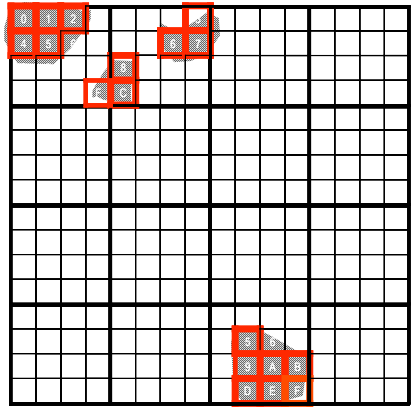

**Figure 3: Example field configuration. The filled regions indicate the extent of the targets to be observed. The grid represents a regular spaxel sampling pattern with red indicating the optimum choice of spaxels to be routed to the spectrograph slit.**

Initially, we consider remapping as applied to the case of one-way multiplexing, where $N_D = 1$ before broadening the discussion to two-way multiplexing, where $N_D > 1$.

In all cases, although the remapping may be arbitrary, it is known, either by design or through post-manufacture calibration. Thus the operator can select the desired pattern by choosing which spaxels are enabled within each cell and downloading appropriate commands to the switcher

In the examples shown $N_S = N_C = 16$ so $N = 256$ and $M = 16$. In the figures, spaxels are coloured by cell and labelled within each cell by a hexadecimal digit for convenience and different cells coloured where necessary to distinguish them. The yellow-hashing indicates the distribution of target spaxels which the system is trying to address.

We need to be clear about the definition of a cell. As defined previously, this is a group of adjacent spaxels of a specified size ($N_C$). However the term cell can refer to such a group at either the input or the output of the input fibre bundle without any implication that they are the same physical fibres.

**Direct mapping:** regular one-to-one correspondence between input and output so that the two patterns are identical (Figure 4). Contiguous groups of spaxels will not be routed to the slit unless they span the boundary between different cells of the Spatial Light Modulator array. The best that can be done is if the field can be repositioned to put one contiguous group on the boundary between 4 cells. The maximum field repositioning required is equivalent to one cell or $\sqrt{N_C}$ in both directions or $N_C = N/M$ choices. This is clearly an inefficient solution. The example shows a success rate of only 5/18 compared to a theoretical maximum of 16/18.

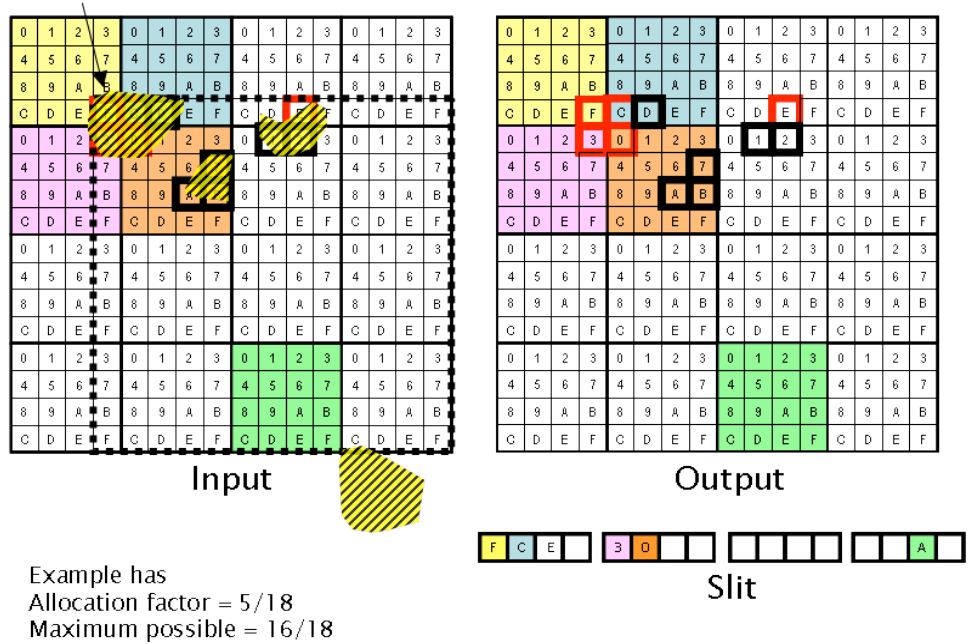

**Figure 4: Illustration of direct mapping. The field is divided into cells of 16 in this example (some are colour-coded for convenience). One of each cell may be selected by the operator to be routed to the slit. The output distribution shows spaxel s at the output of the fibre reformatter before further reformatting to the pseudo slit. Heavy black indicates target spaxels which cannot be routed to the slit. Note that the field has been shifted to maximise the allocation factor by exploiting the connectivity of the groups.**

**Random mapping:** the mapping is randomised so that there is a finite probability that spaxels which are adjacent in the field will be routed to different cells and so have the potential to be routed to the slit (Figure 5). In this example, the allocation factor is optimal.

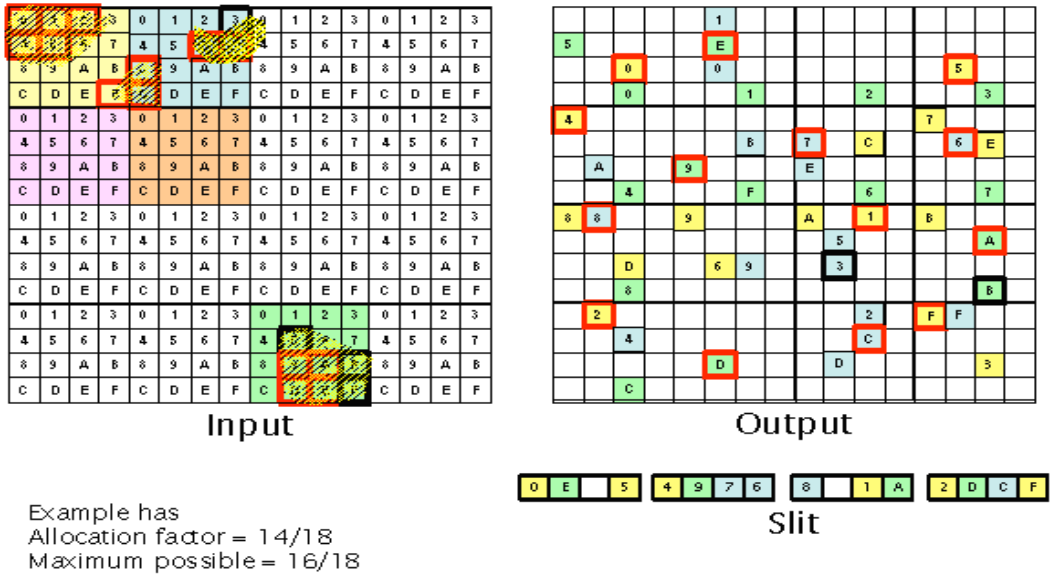

Example has
Allocation factor = 14/18
Maximum possible = 16/18

**Figure 5:** Illustration of random mapping in which each spaxel in a cell is mapped randomly to an output group. Selected groups are identifiable by colour and the spaxel within each cell routed to the slit by its hexadecimal digit.

**Ordered mapping:** each spaxel in an input cell is routed to a different output cell. Thus it can be ensured that groups of contiguous spaxels will be routed to different output cells and hence to the slit. This solution is optimal for creating constellations containing groups of contiguous spaxels of size similar to that of the cell. In this example **(Figure 5)**, the allocation factor is optimal.

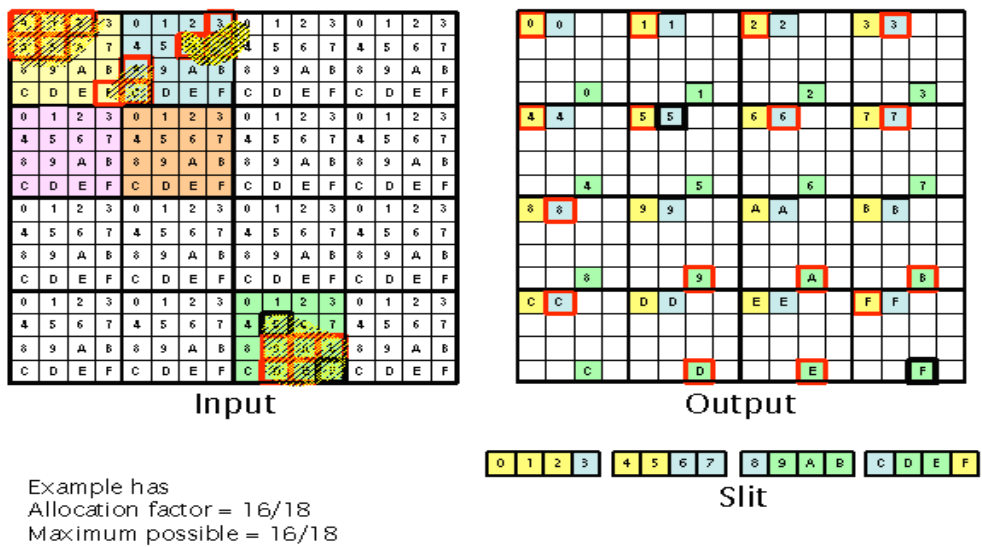

Example has
Allocation factor = 16/18
Maximum possible = 16/18

**Figure 6.:** Illustration of ordered mapping. Here the success rate is near-optimum

## 3.2 Two-way multiplexing

The utility of a two-way multiplexed switcher system is illustrated in Figure 7. Let us first assume that the field is broken into $N_C = 9$ cells with $N_S = 64$ and $N_D = 8$. Therefore $N = 576$ and $M = 72$.

Here the case of a contiguous RoI of similar size to a cell or tile served by one multiplexer is examined. If the mapping is coherent the success rate is roughly $N_S N'_C / Q$ where $Q$ is the number of RoIs and $N'_C$ is the number of cells containing significant numbers of RoIs. If the mapping is coherent this becomes $N'_C / Q = 1$ if $Q = M$. So incoherent mapping will always be preferable unless every tile/cell is substantially populated by RoIs. In the example shown the success rates are 40% and 60% for $Q = 120$ for coherent and incoherent mapping respectively. Also shown is the equivalent case for a one-way multiplexer. In this case, incoherent mapping is mandatory since coherent mapping produced an even worse success rate depending on the contiguity of RoIs. For the example shown, it is less than 25%.

Alternatively we can use this configuration in a different way to illustrate the effect of tiling. In this case, there are 9 tiles, each containing a single cell, so $N_T = 9$ and $N_C = 1$ with $N_S = 64$ and $N_D = 8$ as before giving the same numbers of inputs and outputs as before. The difference from the single-tile interpretation is that remapping can only be done over the extent of a single tile. The consequence of this is to reduce the success rate to $N_D/N_S$ in each tile so the overall success rate is $N'_C N_D/Q$ which, in this example *($N'_C \approx 8$), is ~25%.*

Beware however that the concept of success here does not refer (as in the study of the use of one-way multiplexers) to the definition of an arbitrary selection of spaxels but to the success with which spaxels within a particular contiguous group can be addressed without stipulating precisely which spaxels are successfully routed to the slit. This discussion simply serves to illustrate how such a system might be operated for the special case where most RoIs are in one large group.

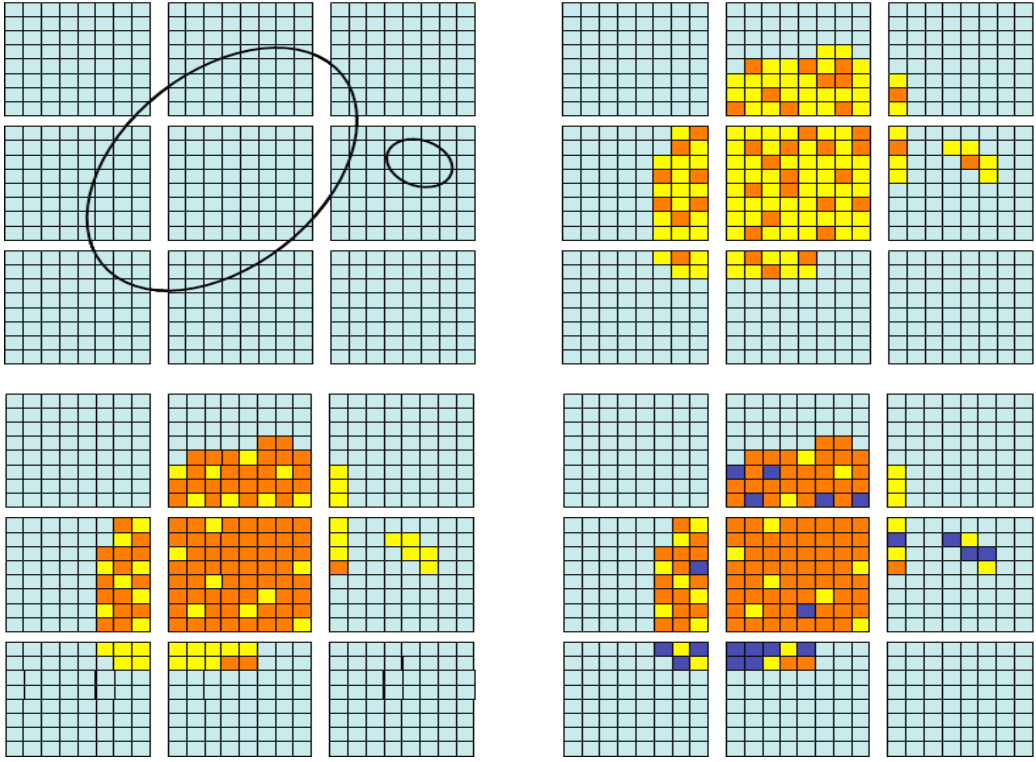

**Figure 7: Illustration of use of a two-way multiplex switcher. In this example there are 9 cells of 64 spaxels of which 8 may be selected. Top-left shows the input field with regions of interest defined as internal to the two ellipses. With incoherent mapping the result at top-right is obtained with yellow marking successful assignments if an input spaxel is routed to the spectrograph slit and orange a failure. If coherent mapping is used the result (bottom-left) is a much lower fraction of successful assignments. For one-way multiplex switching (using a cell equal to one row of each of the 9 tiles and incoherent mapping), the result is even worse – dark blue marks additional unassigned spaxels**

# 4 Performance evaluation

## 4.1 Simulation procedure

The utility of these mapping strategies has been studied for different assumptions about how the observer may define a constellation of RoI. We consider two extremes; a uniform, random selection and a highly clustered distribution which includes a large proportion of RoIs which are in contiguous groups.

Initially, we consider the case of one-way multiplexing and then broaden the analysis to include two-way multiplexing.

If the distribution of RoIs is uniform, the probability of obtaining a given distribution of $Q$ RoIs is calculated from the Poisson distribution that a cell is not unoccupied, normalised to unit probability for $Q = 1$:

$$p_Q(Q) = \frac{N_C}{Q}\left(1 - \exp\left[-\frac{Q}{N_C}\right]\right) \qquad (1)$$

For the case where $Q = M = N_C$, the probability is 63%.

However, if the distribution is non-uniform, the situation must be studied by numerical simulation. A convenient quantification of the distributions is given by the two-point correlation function (e.g. Baugh 2006):

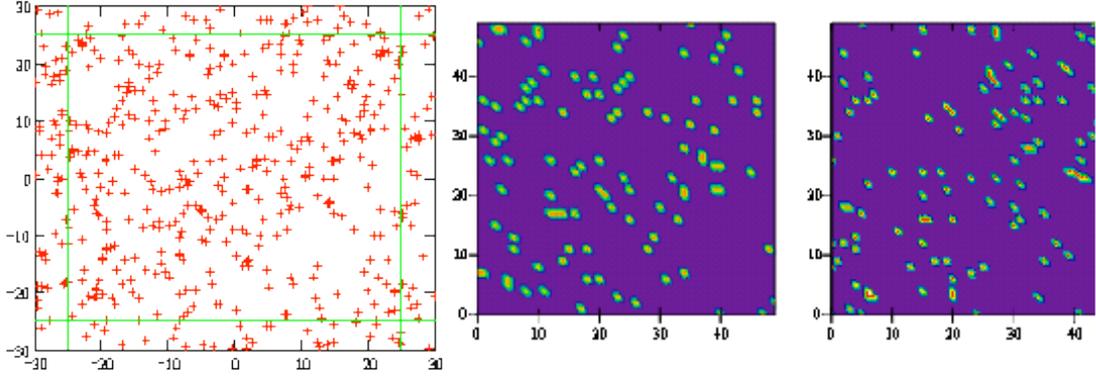

**Figure 8: Result of simulation for uniform distribution with random reformatting. Left: distribution of points from the Monte Carlo generator. Centre, the Boolean image of the original field defined by the green lines. Right: the same after reformatting.**

$$\xi(R) = (R/R_0)^{-\gamma}$$

Where $R$ is the separation of the pairs and $\gamma$ and $R_0$ are fitting parameters. From this the number of pairs with separation $R$ is

$$dN(R) = \rho_0 (1 + \xi(R)) dV_1 dV_2$$

Where $\rho_0$ is the number density of points and $dV_j$ is the search volume at each location. Projecting this onto a plane yields the angular two-point correlation function expressed in terms of the projected separation, $r$,

$$w(r) = 2\int_r^\infty \xi(R)(R^2 - r^2)^{-1/2} dR$$

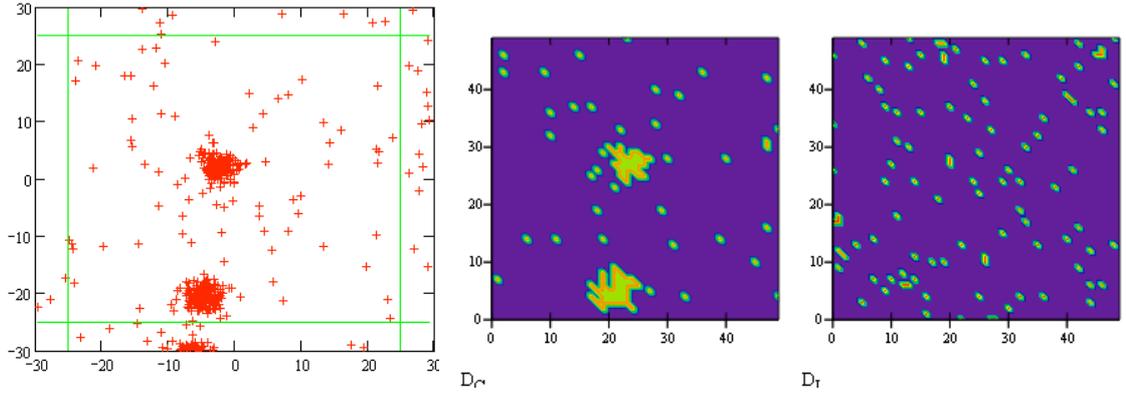

- **Figure 9: Same as previous figure but with $\gamma = 4$. Most of the RoIs are in two clumps of contiguous spaxels which are traced by the red contour in the centre panel.**

The RoI positions are determined by generating a large number of pairs with the correct probability distribution using a Monte Carlo technique. These are used to generate the RoI positions as offsets from one of a much smaller number of "cluster" centres chosen at random from a list of positions determined using the same technique but with different parameters, $\gamma_C$ and $R_{0C}$. A minimum length scale $R_{min}$ is also used to limit the surface density of RoIs on scales below the spaxel size. All these fitting parameters have been adjusted subjectively to give representative types of distribution. However this analysis can be done objectively following this methodology subject to the limitations imposed by neglecting the higher orders of correlation represented by *n*-point correlation functions where *n* > 2.

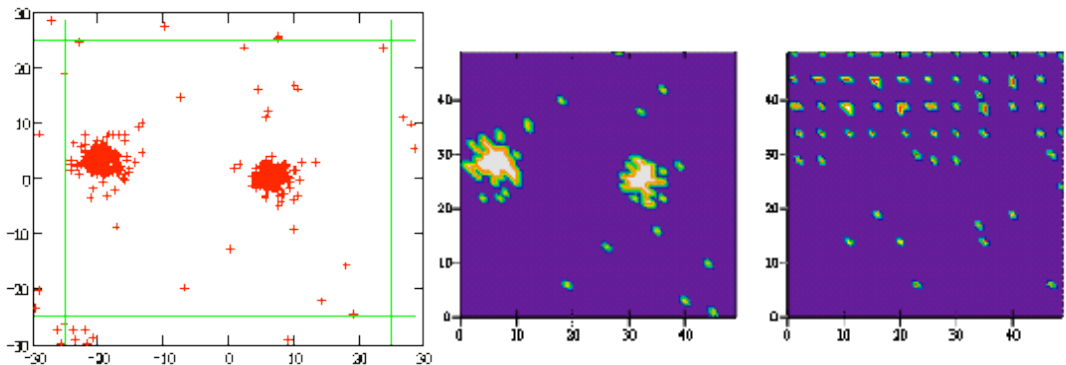

**Figure 10: Same as previous figure but for ordered mapping.**

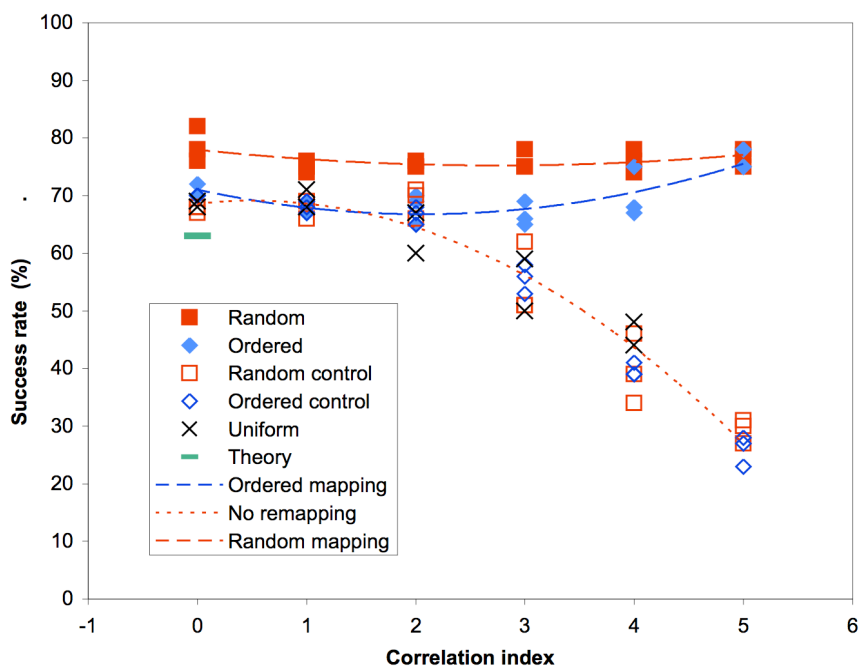

**Figure 11: Plot of success rate (i.e. occupancy) against correlation index for different types of remapping and without remapping ("control"). Datapoints for different realisations are shown together with trend lines (dashed curves).**

The observation is simulated as follows. (a) The distribution of points is generated on a 2-D grid representing the field of view as described above. (b) The distribution of occupied spaxels is determined by binning the points in unit intervals to generate an array with unit value for spaxels with one or more occupant and zero otherwise. The result is a Boolean image defining the desired RoI configuration to simulate how the observer chooses a constellation by interaction with a direct image of the field. (c) This image is then re-mapped in one of the ways already described to generate a new Boolean image in which the RoIs appear in different locations. This simulates the distribution input to the switcher. (d) The re-ordered Boolean image is then rebinned in square cells of $N_S$ units to yield a rebinned image of $M$ blocks containing the occupancy of that pixel. The occupancy represents the number of competing RoIs in the original Boolean image defining the desired constellation of RoIs. However, only one of these may be routed to the output so that its spectrum may be recorded. (e)

The number of RoIs that can be successfully routed is obtained by counting the number of blocks with non-zero occupancy. (f) This procedure is repeated with the field centre at different spaxel positions to find the optimum solution.

### 4.2 Simulations for one-way multiplexing

Simulations were made with $N = 2500$ and $M = 100$. Although this format is appropriate for biomedical applications, it is too small for an astronomical implementation. However the same principles apply regardless of scale.

Results of individual simulations with different random number seeds are shown in Figure 8 for a uniform distribution ($\gamma = 0$) and in Figure 9 for a highly-concentrated distribution ($\gamma = 4$) in which the RoIs are grouped into a small number of contiguous regions. These are for random mapping. Figure 10 shows a simulation with the same simulation parameters, but for ordered mapping. These were done with $\gamma_C = 1.5$ to give a realistic distribution but, as a control, some were done with $\gamma = \gamma_C = 0$ without remapping (labelled "uniform" in the figure).

The success rate for the different remapping options as a function of correlation index is shown in Figure 11. The success rate for uniform distributions is consistent with the theoretical prediction of Equation 1 (marked as a bar) supporting the validity of the simulation.

As expected the advantage of remapping is negligible for the unclustered distribution (~70% of 100 possible RoIs are observable in both cases), but that there is a big difference for a clustered distribution: without remapping the success rate drops to ~25% but the remapped case increases to 75% with the highest value for the random mapping, although the advantage becomes less marked for the highest degrees of clustering.

### 4.3 Two-way multiplexing

The simulations described above were repeated for the same number of input and output spaxels but with an output multiplex value of $N_D = 4$ (25) implying $N_C = 25(4)$ and $N_S = 100(625)$. The results are shown in Figure 11

The previous results for $N_D = 1$ and random mapping are shown for reference. As the multiplex factor increases the success rate increases to 100% for $N_D = 25$. Without remapping, the success rate is reduced for highly clustered RoIs but is also increased as the multiplex factor is increases. This is expected since increasing the multiplex means that there is an increased chance of spaxels which are adjacent on the sky being independently routed to the output without needing incoherent mapping, but this process is enhanced when large groups of contiguous spaxels are required by the incoherent mapping.

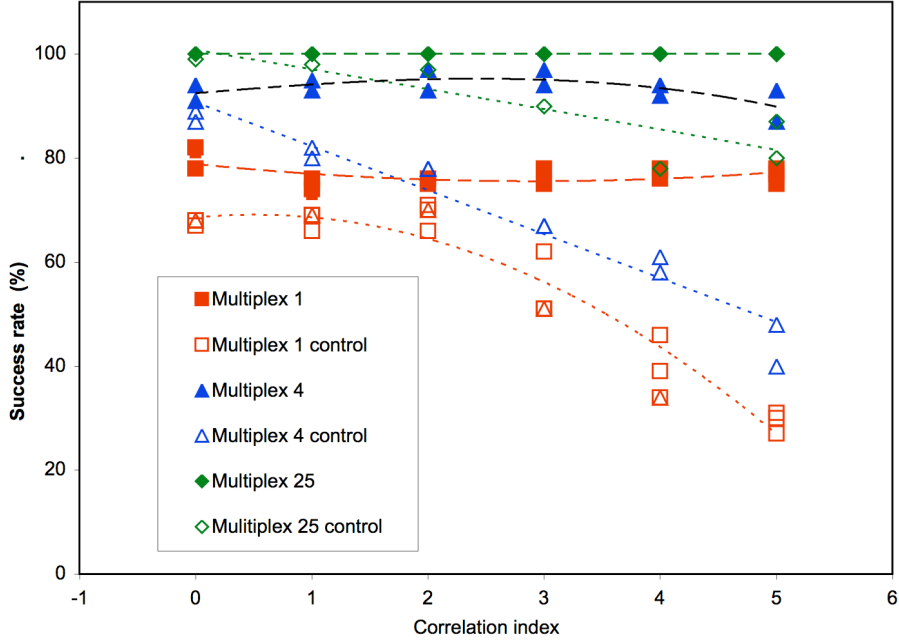

Figure 12: Same as Figure 11 but showing results for two-way multiplexing by "multiplex" factor 4 or 25. Only random mapping is used for the incoherent case. Results for 1-way multiplex are labelled as "Multiplex 1". The term "control" refers to coherent mapping.

### 4.4 Comparison with other methods

It is also possible to estimate the success rate when using a single large Integral Field Unit (sIFS option) or a multiplicity of mini-IFUs which can be positioned anywhere in the field (mIFS option). This was done by binning the Boolean image of the field on a variety of scales and shapes and labelling each spaxel in the original unbinned image by the occupancy obtained if it was a maximum for that bin, i.e. equal to the number of spaxels in that bin. This was done for all the dithered images, offset by integral numbers of spaxels to ensure that the calculation of contiguity was not influenced by proximity to a bin edge. The dithered images were combined so that each spaxel retained the maximum occupancy assigned to it in any of the images. This generated a map of contiguous spaxels in which each spaxel is labelled by the contiguity factor which is the number of contiguous spaxels in the group in which it resides (if any). By forming a histogram of the contiguity factor it is possible to estimate the success rate in placing mini-IFUs at the optimum location to address the largest contiguous groups, then to address the next largest groups until the number of mini-IFUs was exhausted. From this, the number of spaxels which could be addressed by such a method can be estimated.

For the single-IFU configuration we assumed $M = 100$ spaxels.. For the *mini-IFU* configuration we assumed 10 mini-IFUs each with 10 spaxels to produce the same number of outputs as for the monolithic approach, $M = 100$. Similarly, the format is too small for astronomy but the results are scalable to a much larger systems – e.g. 50 mini-IFUs each with 200 spaxels giving $M= 10^4$

This procedure is likely to overestimate the success rate because: (a) it does not take into account the possibility of the fields of different mini-IFUs overlapping and (b) it does not account for a zone of avoidance around each IFU. The latter effect can be estimated by assuming that the inter-IFU gap is equal to 10% of the IFU field diameter so the effective success rate is reduced by 20%. The former effect may be partly compensated by the fact that if a mIFU has too few spaxels to address a complete group, the excess spaxels are not reassigned to another IFU. This is reasonable since this would otherwise make an IFU field overlap likely. However, a partial correction is made by calculating the probability that the remaining unassigned spaxels in all the mIFUs combined will be assigned to the remaining unassigned RoIs in the field if the distribution of RoIs is uniform. Thus we multiply the contiguity factor by 0.8 to allow for the zone of avoidance but make no other correction.

The success rate for the sIFS option with the same $M$ was estimated by locating the region with the highest contiguity factor and counting the number of RoIs in the Boolean image which are within a square contiguous field containing $M$ spaxels centred at that location. This procedure maximises the number of RoIs which can be addressed by the IFU.

Finally we simulated the case for pure Multiobject spectroscopy (MOS option) where the spaxels are independently positionable but isolated from each other by the encapsulation of the fibre and the limitations of the deployment mechanism. In this case we assume that each spaxel is packaged such that there is a buffer width equivalent to one spaxel around each spaxel, so that the maximum packing fraction is 1/9. This can be simulated using the above methodology by adopting a binning factor of 3 (i.e. $N_C = 90$) while keeping the number of RoIs the same as before, $M = 100$.

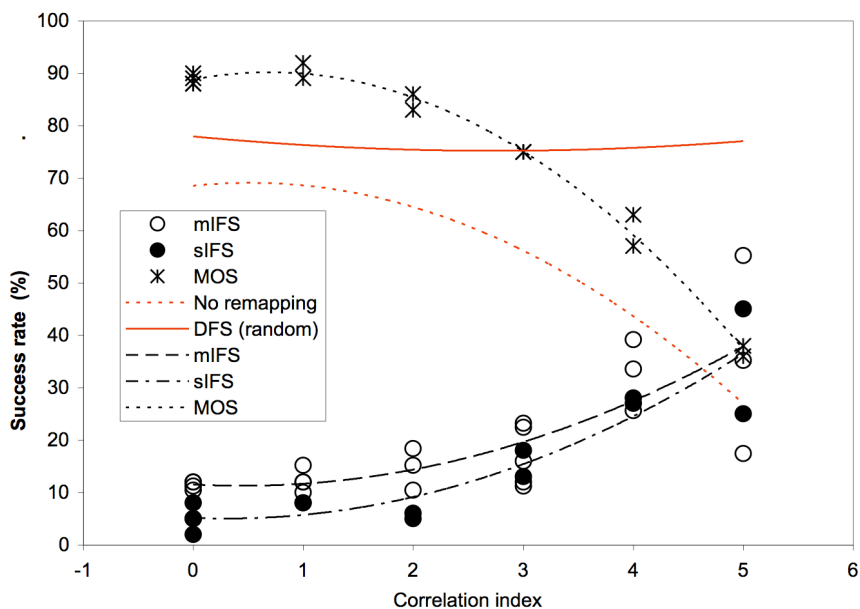

**Figure 13: As Figure 11 but for the non-Diverse Field Spectroscopy options discussed in the text. For comparison the trendlines for Diverse Field Spectroscopy with random mapping and direct mapping are also shown.**

The results of these simulations are shown in Figure 13, which also indicates the run of results for Diverse Field Spectroscopy with random mapping. Both IFS options are very inefficient for unclustered distributions but increase in success with clustering as expected. However the success rate is extremely variable for the highly clustered distributions and always less than the remapped Diverse Field Spectroscopy options. The MOS option is very efficient for unclustered distributions since each RoI can be addressed by a single spaxel except when the surface density of targets is too high. However it loses efficiency for clustered distributions, producing results similar to Diverse Field Spectroscopy without remapping or the IFS options. Although the MOS results make it an attractive option for unclustered distributions, it must be remembered that it totally excludes the possibility of observing RoIs in any sort of contiguous group unlike all the other options.

# 5  Application to the real world

## 5.1  Simulations with the real sky

For a reality check, the technique was applied to a selection of real astronomical objects for a switch format that is commercially available for single mode fibres.

We adopted a cell format for a switch currently available commercially for single-mode fibres manufactured by CrossFiber Inc. The format is an array of 48 x 48 spaxels arranged into $N_C$ = 18 cells each of $N_S$ = 8 x 16 = 128 spaxels of which $N_D$ may be routed to the output. Thus $N$ = 2304 and $M$ = 198 so the downselection factor is $G$ = 11.6. In the simulations, $N_D$ was altered to explore the success rate as the down-selection factor was varied.

The targets were taken from the Messier (1781) catalogue including both galaxies and star clusters. As such it provides an illustrious and illustrative selection of different types of association, clustering and contiguity with which to verify the purely synthetic catalogues studied above. The SAURON observation of Lyman-alpha emitter LAB-1 (Paper II, Weijmann's et al. 2009) was also analysed in the same way.

Images[†] of a number of Messier catalogue members were selected at random. These were binned into 48 x 48 spaxels containing unit value if the intensity exceeded a specified threshold or zero otherwise. The 100 brightest spaxels were identified. These Boolean images were then binned into 18 cells and the population, $n$, of identified spaxels in each cell determined. The success rate was calculated as the average of $n/N_D$

Table 1 gives the success rate for the design value, $N_D \approx 11$ and also the value of $N_D$ required a specified success rate. Without random mapping, the values show a large scatter ranging from $N_D \approx 14$ for fields with points distributed at random such as an open star cluster M52 to $N_D \approx 65$ for a heavily concentrated distribution of ROIs such as M84.

---

[†] http://hou.lbl.gov/vhoette/Explorations/MessierGallery/index.html.

Table 1: Success rate for $N_D$ =11 with 100 ROIs, and the value of $N_D$ required to obtain the specified success rate for randomly-selected objects from the Messier catalogue and LAB-1.

| Image | Rank | Success rate (%) for $N_D$=11 | | Original mapping $N_D$ which gives success rate = | | | | | | Random mapping $N_D$ which gives success rate = | | | | | |
|---|---|---|---|---|---|---|---|---|---|---|---|---|---|---|---|
| | | Original | Random | 50% | 60% | 70% | 80% | 90% | 100% | 50% | 60% | 70% | 80% | 90% | 100% |
| M84 | 1 | 22 | 97 | 31 | 38 | 46 | 53 | 52 | 65 | 4 | 5 | 6 | 7 | 9 | 13 |
| M95 | 2 | 28 | 99 | 27 | 35 | 44 | 53 | 61 | 70 | 4 | 5 | 6 | 7 | 9 | 12 |
| M109 | 3 | 29 | 100 | 26 | 33 | 41 | 48 | 55 | 63 | 4 | 5 | 6 | 7 | 8 | 11 |
| M2 | 4 | 31 | 99 | 24 | 32 | 39 | 47 | 55 | 70 | 4 | 5 | 6 | 7 | 8 | 12 |
| M90 | 5 | 28 | 99 | 22 | 30 | 38 | 46 | 54 | 65 | 4 | 5 | 6 | 7 | 9 | 12 |
| M97 | 6 | 32 | 98 | 19 | 24 | 30 | 40 | 54 | 68 | 4 | 5 | 6 | 7 | 8 | 13 |
| M61 | 7 | 31 | 100 | 19 | 23 | 29 | 40 | 52 | 65 | 4 | 5 | 5 | 6 | 8 | 11 |
| M12 | 8 | 57 | 100 | 9 | 13 | 18 | 25 | 39 | 53 | 4 | 5 | 5 | 6 | 7 | 10 |
| M22 | 9 | 70 | 98 | 6 | 8 | 11 | 16 | 23 | 31 | 4 | 5 | 6 | 7 | 8 | 12 |
| M52 | 10 | 96 | 99 | 4 | 5 | 6 | 7 | 9 | 14 | 4 | 5 | 6 | 7 | 9 | 13 |
| LAB-1 | - | 43 | 98 | 12 | 15 | 25 | 37 | 50 | | 4 | 5 | 6 | 7 | 9 | 12 |

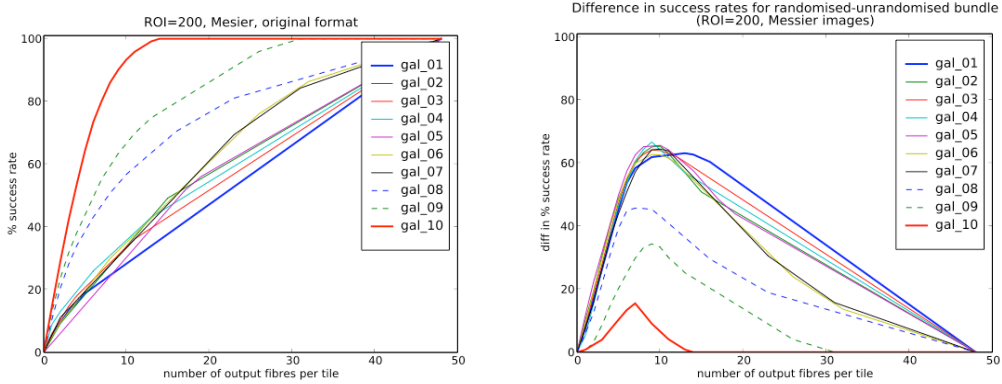

Figure 14: Success rate as a function of the number of outputs per cell, $N_D$ for the 10 images of Messier objects. Left – original unrandomised mapping Right – improvement with randomised mapping.

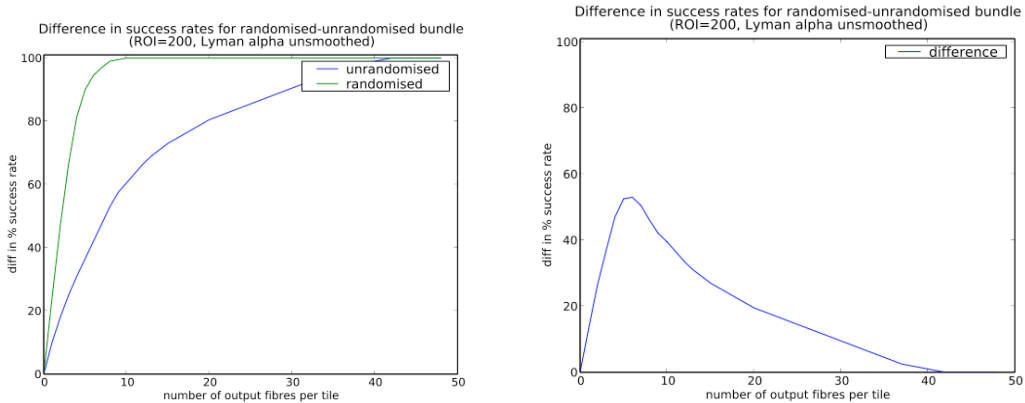

Figure 15: Same as previous figure but for LAB-1 (both original and remapped are shown on the left).

With random mapping, all the fields show similar results with $N_D$ = 10-13 being optimal (i.e. giving the greatest improvement when random remapping is used).

Figure 14 and Figure 15 shows the success rate as a function of $N_D$ for the unrandomised case and the improvement in success rate with the randomised mapping. The improvement is greatest for values of $7 < N_D < 25$ which is close to the value needed to maximise the success rate with random mapping in an absolute sense and similar to the number of outputs provided in the commercial device.

Multiple realisations were used to estimate the statistical uncertainty. From this we infer that the typical RMS scatter in the abscissa is ~5%.

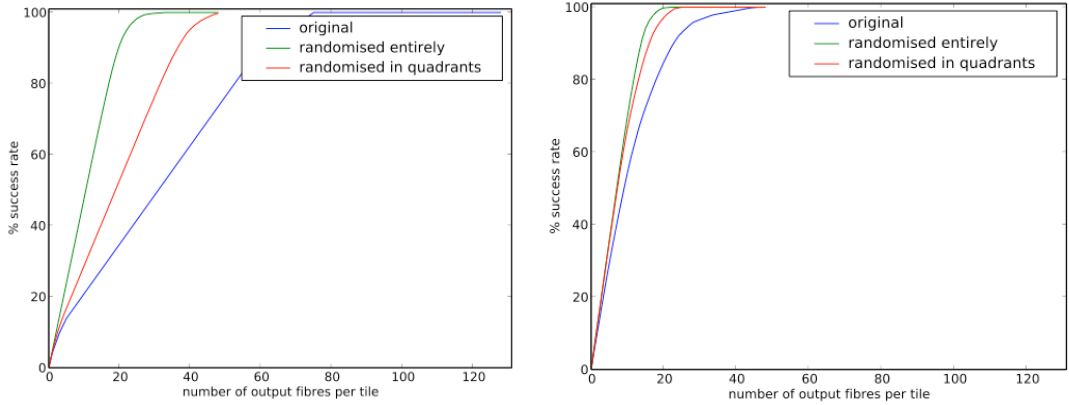

**Figure 16: Effect or randomising in quadrants for two of the Messier objects studied.**

We also tested the effect of the tile size in limiting scope for randomisation by dividing the images into 4 equal quadrants and randomising only within each quadrant. Figure 16 shows that this requires a higher number of outputs to achieve a given success rate depending on the amount of clustering of ROIs already present.

Finally the same analysis was performed with artificial catalogues of the type described in §3. The results are shown Figure 17. The curves are labelled by the correlation index used to generate the distribution of ROIs. Here the optimum number of outputs is $4 < N_D < 10$, roughly half that required by the Messier catalogue. This is consistent with the LAB-1 image for which the optimum is $N_D \approx 7$.

The difference may be because the objects in the Messier catalogue (or, more precisely, the distribution of ROIs tracing the light distribution) are more clustered than in the artificial catalogues. This is supported by the overall lower benefit from randomisation compared with the Messier catalogue.

The conclusion of this analysis is that, depending on the reality of the RoI distribution, random mapping is likely to increase success rates in assigning spaxels to arbitrary selections of RoIs by up to 80% depending on the clumpiness of the RoI distribution, with negligible benefit from completely unclumped distributions. For the configuration described here the optimum number of outputs/cell is 5-15.

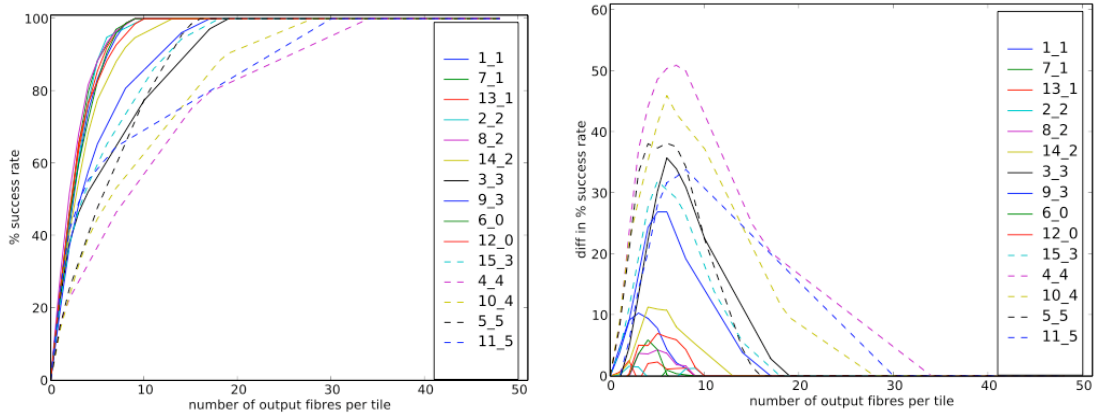

Figure 17: Success rate as a function of the number of outputs per cell, $N_D$ for artificial catalogues as described in the text. Left – original unrandomised mapping Right – improvement with randomised mapping. The curves are labelled by the run number *n* and the correlation index as "*n_γ*" with high values of γ indicating greater clustering.

# 6  Conclusions

We have continued the exploration of methods to provide efficient and affordable spectroscopic instrumentation on Extremely Large Telescopes with a study of techniques for highly-multiplexed spectroscopy; both the strategies and their implementation. This paper deals specifically with the advantage to be gained by adopting a random mapping of the fibre bundle.

In the previous paper, we made a case that cosmological studies in particular are likely to require a mixture of multiple object and integral field spectroscopy to allow the field to be addressed in an arbitrary pattern of spaxels which may be isolated or in contiguous groups: Diverse Field Spectroscopy. Mindful of the limitations of mechanical field-selector deployment in delivering sufficient stability, we have examined fibre-based options without large-scale mechanisms to down-select from a fibre system covering a large field of view to a smaller number of fibres which can be fed into a number of spectrographs. These may use fibre-switching networks already available commercially. We believe that only through the industrialisation of the switching systems, and the adoption of production engineering techniques widely used elewhere, will it be possible to provide the desired high multiplex factors while maintaining good operational efficiency.

Such solutions generally require an incoherent mapping between the sky and the multiplex switcher, so that spaxels which are adjacent on the sky can be independently routed to the spectrographs as required by Diverse Field Spectroscopy. We examine different mapping strategies and make detailed simulations of their performance for different degrees of clustering of the regions of interest to be addressed. The simulations use artificial and real catalogues of objects to define regions of interest and calculate the success rate in targeting as a function of degree of clustering and the geometry of the switching system used, especially the number of outputs per cell.

Using artificial catalogues we show the advantage of the remapping for clustered distributions and the superiority of the Diverse Field Spectroscopy approach over single large-field IFUs and multiple single fibres or small groups in such cases.

Using real and artificial catalogues, we demonstrate the advantage in remapping and quantify the optimal down-selection factors. Depending on how clustered the distribution of ROIs is before randomisation, increases in success rate of up to 80% may be obtained for a down-selection factor of ~10, although a range 5 – 20 would be optimal for different distributions. Devices of this type are currently available for single-mode fibres.

Although the ideal solution is a huge integral field unit covering the entire field, for reasons explained in Paper I, this is not feasible within a realistic budget. A practical system must sample selected regions of the field. We have demonstrated the best way to do this, giving observers maximum flexibility in their choice of targets.

## Acknowledgements

We thank Miles Padgett, John Girkin and Gordon Love for their work on the bio-medical IFS applications which inspired some of this work. In particular, we are grateful to Miles for proposing the mapping scheme.